\documentclass[aps,prb,superscriptaddress,twocolumn]{revtex4-2}
\usepackage{graphicx}
\usepackage{dcolumn}
\usepackage{bm}
\usepackage{amssymb}
\usepackage{amsmath}
\usepackage{hyperref}
\usepackage{epsf}
\usepackage{xcolor}

\def\kMn{$\kappa$-(BETS)$_2$\-Mn[N(CN)$_{2}$]$_3$}
\def\kCl{$\kappa$-(BE\-DT\--TTF)$_2$\-Cu\-[N\-(CN)$_{2}$]Cl}

\def\kCN{$\kappa$-(BE\-DT\--TTF)$_2$\-Cu$_2$(CN)$_{3}$}

\def\TMI{$T_{\rm MI}$}
\draft

\begin{document}
\preprint{ESR kappa-Mn}

\title{ESR Investigations of the Magnetic Anisotropy in $\kappa$-(BETS)$_2$Mn[N(CN)$_{2}$]$_3$ }

\author{Zhijie Huang}
\affiliation{1.~Physikalisches Institut, Universit\"at Stuttgart, Pfaffenwaldring 57,
70569 Stuttgart, Germany}
\author{Marvin Schmidt}
\affiliation{1.~Physikalisches Institut, Universit\"at Stuttgart, Pfaffenwaldring 57,
70569 Stuttgart, Germany}
\author{Savita Priya}
\affiliation{1.~Physikalisches Institut, Universit\"at Stuttgart, Pfaffenwaldring 57,
70569 Stuttgart, Germany}
\author{Mark Kartsovnik}
\affiliation{Dresden High Magnetic Field Laboratory (HLD-EMFL), Helmholtz-Zentrum Dresden-Rossendorf, D-01328 Dresden, Germany}
 \author{Natalia Kushch}
\affiliation{Federal Research Center of Problems of Chemical Physics and Medical Chemistry,
Russian Academy of Sciences, 142432 Chernogolovka, Russia}
 \author{Martin Dressel}
\affiliation{1.~Physikalisches Institut, Universit\"at Stuttgart, Pfaffenwaldring 57,
70569 Stuttgart, Germany}

\begin{abstract}
The two-dimensional molecular conductor $\kappa$-(BETS)$_2$Mn[N(CN)$_2$]$_3$ has been studied
because of the intriguing magnetic coupling of the molecular $\pi$-electrons to the Mn$^{2+}$ ions.
Utilizing X-band electron spin resonance spectroscopy  we have performed comprehensive investigations of the magnetic properties, in particular on the temperature and angular dependences of the spin susceptibility, the $g$-factor and the linewidth. Due to the $\pi$-$d$-coupling,  a rearrangement of the $\pi$-spins occurs:
At low temperatures the $g$-factor shifts enormously with a pronounced in-plane anisotropy that flips as the temperature decreases; the lines broaden significantly; and the spin susceptibility increases upon cooling with a kink at the phase transition. By carefully analyzing the angular dependence of $g(\theta)$ and $\Delta H(\theta)$
we reveal the influence of anisotropic Zeeman interaction in addition to spin-phonon coupling. We conclude the presence of two magnetically distinct BETS chains
and discuss the possibility of altermagnetic order.
\end{abstract}



\date{\today}%
\maketitle


\section{Introduction}
\label{sec:introduction}
The BEDT-TTF radical cation salts and the related BETS family
[BEDT-TTF stands for bis(ethylene\-di\-thio)tetra\-thia\-fulvalene while BETS means bis(ethylene\-di\-thio)tetra\-selena\-fulvalene]
are renowned for their exciting electronic properties, such as superconductivity or Mott metal-insulator transition \cite{ToyotaBook,Dressel20}.
In some systems, one additionally finds a subtle interplay of itinerant $\pi$-electrons in the conducting layers of organic molecules and localized $d$-electrons in the insulating anion layers leading to remarkable phenomena
such as magnetic ordering or field-induced superconductivity \cite{Uji01,Balicas01,Uji03,Kobayashi04,Enoki04,Coronado04,Blundell04}.
Here we focus on the strongly correlated molecular conductor \kMn\ where the Mn$^{2+}$ ions possess $S=5/2$ spins
\cite{Kushch08,Zverev10,Vyaselev11b,Vyaselev11a,Vyaselev12a,Vyaselev12c,
Kartsovnik17,Kushch17,Vyaselev17,Zverev19,Riedl21,Thomas24}.

\begin{figure}[h!!]
    \centering
        \includegraphics[width=\columnwidth]{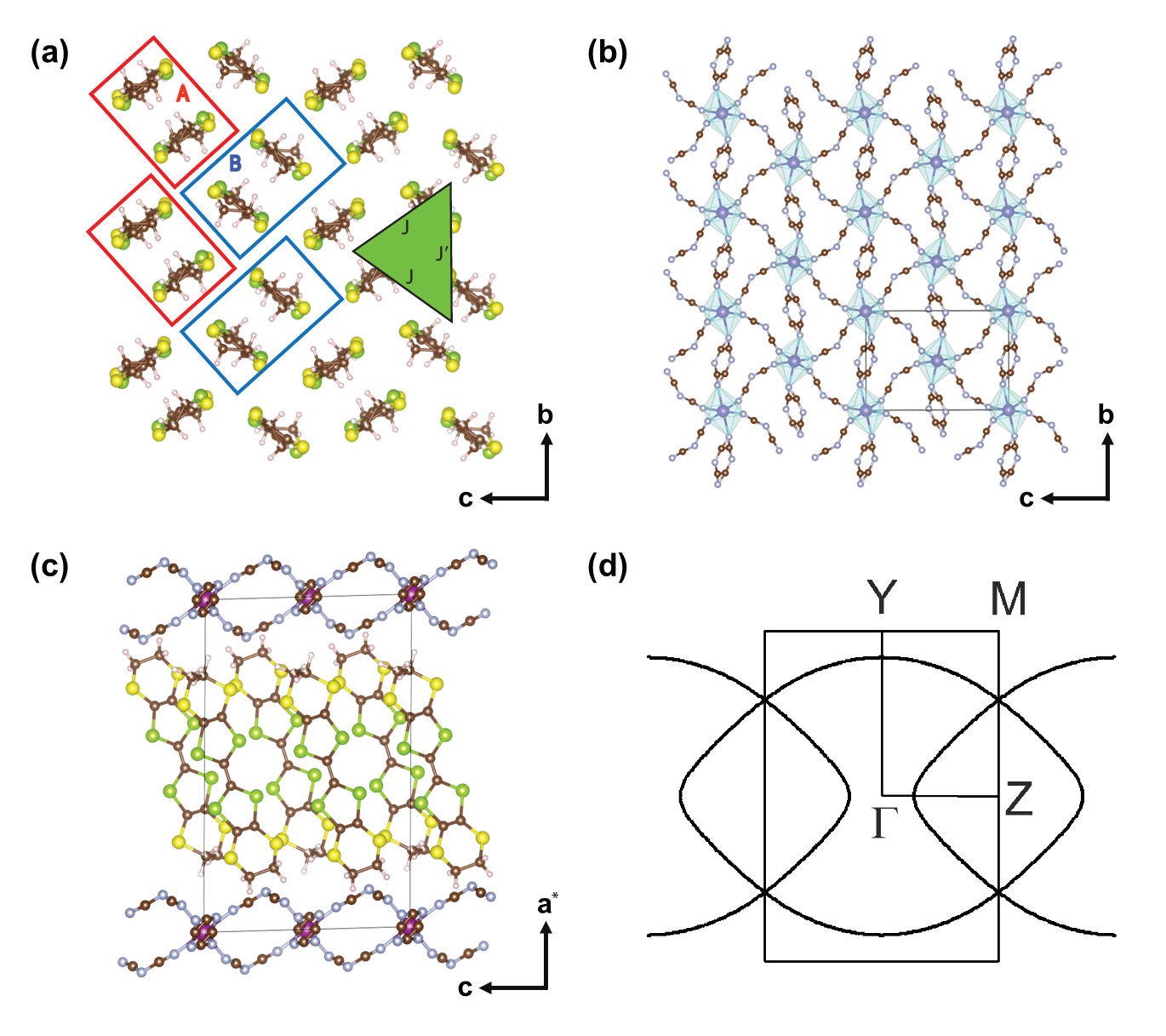}
        \caption{The two-dimensional charge-transfer salt \kMn\ crystallizes in a monoclinic unit cell containing two formula units ($Z=2$); the space group is $P2_1/c$. (a)~Arrangement of the BETS molecules projected onto the $bc$-plane. The distinct dimers A and B are indicated by red and blue squares. They form a triangle as illustrated in green, with the coupling close to frustration $J^{\prime}/J \approx 1$. (b)~The bulky anions are formed by Mn$^{2+}$ connected to six neighbors via N$\equiv$C-N-C$\equiv$N ligands. (c)~The layered structure becomes obvious when projected onto the $a^*c$-plane. (d)~2D Fermi surface of \kMn\ calculated by the first-principle density-functional theory (DFT
        ) \cite{Zverev19}.}
    \label{fig:structure}
\end{figure}

The quasi-two-dimensional structure consists of alternating BETS donor layers,
separated along the $a^*$-direction by insulating Mn[N(CN)$_2$]$_3^-$ layers \footnote{Conventionally, $a^*\perp(bc)$, but our experimental accuracy does not allow us to distinguish $a$ and $a^*$, which differ by less than $2^{\circ}$.}, as shown in Fig.~\ref{fig:structure}.
The polymeric anionic structure contains one Mn atom at the inversion center and two independent N(CN)$_2$ ligands. Each Mn atom is connected to six neighboring metal atoms via N(CN)$_2$ bridges. \cite{Kushch08,Schlueter04}.
The two types of (BETS)$_2^+$ dimers, labeled A (red squares) and B (blue squares), are distinguished by symmetry as they are slightly tilted against the $a^*$-axis; consequently we can identify two stacks along the $b$-axis, one with dimer A and one with dimer B.

The coupling within the stack ($b$-direction) is stronger than the interaction ($c$-direction) between the stacks \cite{Zverev10}.
The density functional theory (DFT) band structure agrees well with previous calculations using the extended H{\"u}ckel method. Although in both cases correlation effects are neglected, the calculated Fermi surface [Fig.~\ref{fig:structure}(d)] was confirmed by angular magnetoresistance oscillations (AMRO) experiments.
However, the importance of the electron-electron interaction is unambiguously  concluded from studies of high-field magnetoresistance and magnetic quantum oscillations \cite{Zverev10,Kartsovnik17,Zverev19}.

The compound is metallic upon cooling down to about $T_{\rm MI} = 21$~K, then a sharp metal-insulator transition occurs. There is no evidence of structural or charge ordering and no indications of density wave formation  that could cause a gap on the whole Fermi surface.
The metal-insulator transition can be suppressed by moderate pressure of $\sim 0.55$~kbar, with superconductivity observed up to $T_c=5$~K.

From $^1$H and $^{13}$C NMR studies, a long-range staggered structure of the $\pi$-electron spins was inferred, which already occurs at \TMI, while the $3d$ Mn$^{2+}$ spin moments form a disordered, tilted structure.
Only at low temperatures does the $\pi$-$d$ interaction cause a glassy freezing of the manganese spins \cite{Vyaselev12a,Vyaselev12c,Vyaselev17}. Ab-initio calculations confirm the rather weak coupling between the two subsystems \cite{Riedl21}.
Torque measurements identify a field-induced transition at temperatures below \TMI\, which resembles a spin flop
assigned to the spins of the $\pi$-electrons located on the organic BETS molecules \cite{Vyaselev11b}.
Theoretical considerations point to a spin-vortex crystal order due to ring exchange in \kMn\ \cite{Riedl21}, which could also be relevant for the explanation of the gapped ground state in the quantum spin liquid \kCN\ \cite{Riedl19,Miksch21}.

Here we present comprehensive X-band electron spin resonance (ESR) measurements on \kMn\ single crystals at different temperatures and crystallographic directions.
At low temperatures the ESR properties are affected by the Mn$^{2+}$ ions via $\pi$-$d$-coupling and antiferromagnetic arrangement within the $\pi$-spins: The $g$-factor shifts enormously with a pronounced in-plane anisotropy that flips as the temperature decreases; the lines broaden significantly; and the spin susceptibility increases upon cooling with a kink at the phase transition.
From analyzing the angular dependence of the ESR parameters at $T=300$~K
we can reveal anisotropic Zeeman interaction, leading to a doubling of the periodicity of the linewidth when the field is rotated out of plane.
Here the $\pi$-$d$-coupling is of minor relevance.
The observations evidence that the two molecular chains are magnetically distinct.
The complex magnetic structure makes the compound a possible altermagnet.

\section{Materials and Methods}
\label{sec:materialsandmethods}

The single crystals of \kMn\ were grown electrochemically \cite{Kushch08} and had the typical size of $20 \times 850 \times 1200~\mu{\rm m}^3$.
The electron spin resonance (ESR) was measured in a continuous-wave X-band spectrometer (Bruker ESP 300) at 9.5~GHz. The temperature dependence of the ESR properties was recorded down to $T=4$~K by utilizing an Oxford Instruments continuous-flow helium cryostat. These experiments were carried out along different orientations; furthermore, we recorded the angular dependence at certain temperatures by rotating the crystals using a goniometer.

\section{Results and Analysis}
\subsection{Temperature Evolution}
\label{sec:temperaturedependene}

\begin{figure}[h]
    \centering
        \includegraphics[width=0.8\columnwidth]{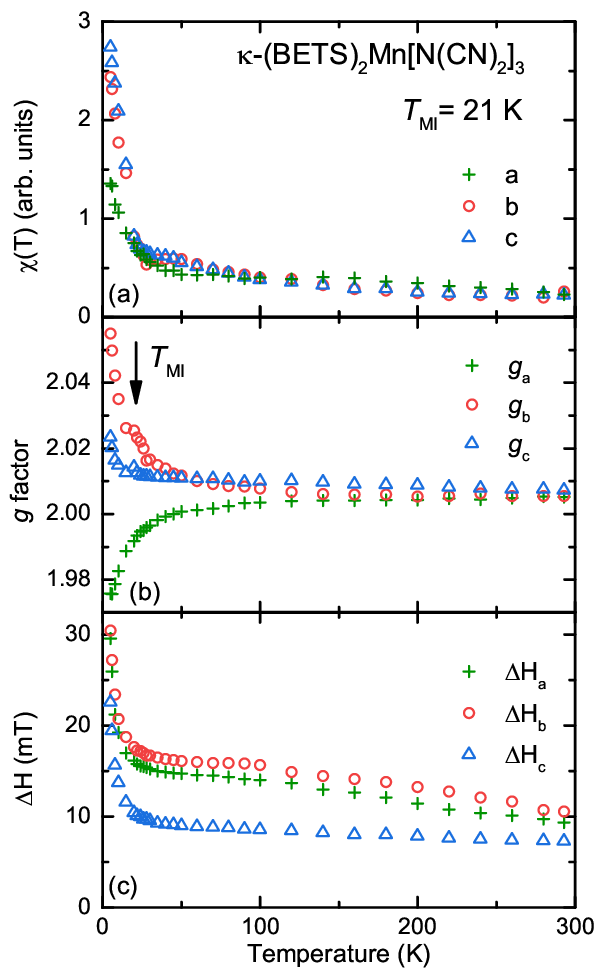}
        \caption{ESR parameters extracted from temperature-dependent X-band measurements on \kMn\ along the three crystallographic directions $H \parallel a^*$, $b$, and $c$, indicated by green pluses, red circles and blue triangles, respectively.  (a)~The spin susceptibility is proportional to the integrated absorption intensity.
        A dramatic increase occurs below $T_{\rm MI}=21$~K. While the experiments within the $bc$-plane are measured on the same crystal, the $a^*$-axis data are taken on a different crystal and had to be normalized by a factor of 3. (b)~The $g$-factor starts to deviate at lower temperatures due to the strong effect of the manganese ions. At $T_{\rm MI}$ anomalies are observed along the $b$- and $c$-axes due to the rearrangement of the $\pi$-spins. (c)~Temperature behavior of the ESR linewidth $\Delta H$.}
    \label{fig:ESR1}
\end{figure}

In order to get an overview on the magnetic behavior of \kMn, the temperature dependence of the ESR parameters obtained by X-band spectroscopy within the $bc$-plane and perpendicular to it is reproduced in Fig.~\ref{fig:ESR1} \cite{Schmidt24}.
In all three directions the spin susceptibility continuously increases as the temperature is reduced; below approximately 25~K it rapidly rises, marking the transition at \TMI.
While the present experiments are carried out at rather low fields of approximately 0.3~T, this phase transition was not detected
by SQUID measurements recorded at 7~T \cite{Kushch08} because in this case mainly the Mn$^{2+}$ spins are probed but not the $\pi$-electron spins.

The temperature evolution of the $g$-factor is plotted in Fig.~\ref{fig:ESR1}(b) for the three crystal directions.
Below $T\approx 100$~K magnetic interactions between the $\pi$-spins and effects of the Mn$^{2+}$ ions become obvious as the signal strongly deviates from the free electron value.
We find that $g(T)$ gradually decreases for the external field oriented out of plane, but strongly rises for the in-plane direction. This becomes even more significant when the transition is approached. For $H\parallel b$ the effect is most pronounced.

In panel (c) of Fig.~\ref{fig:ESR1} the ESR linewidth $\Delta H$ is plotted as a function of temperature. For all directions a pronounced increase is observed upon cooling; $\Delta H_c$ is smallest for the $c$-direction. Again, below 25~K the arranged magnetic moments cause a significant local field and enhanced scattering of the $\pi$-electron spins leads to a rapid rise of $\Delta H(T)$. Proton NMR spectroscopy comes to the same conclusions \cite{Vyaselev11b}.

Compared to other $\kappa$-salts with non-magnetic metal ions in the anion layer, such as
the spin liquid candidates \kCN\ or $\kappa$-(BE\-DT\--TTF)$_2$\-Ag$_2$(CN)$_{3}$, the spin susceptibility and the linewidth of \kMn\ become much larger at low temperatures, evidencing some coupling among the magnetic moments.
Previous reports suggest that the $\pi$-$d$-interaction is rather weak \cite{Riedl21} and negligible at elevated temperatures.
Accordingly our X-band measurements cannot resolve the hyperfine structure of the $S=5/2$ Mn$^{2+}$; it just contributes to the broadening of the lines at low temperatures in the insulating state.
For that reason we suggest Q-band and W-band ESR measurements for a better resolution in order to gain more insight; additionally the hyperfine couplings could be studied applying dedicated spectroscopies, such as electron nuclear double resonance (ENDOR) or hyperfine sublevel correlation spectroscopy (HYSCORE).

\subsection{Angular Dependence}
\label{sec:angulardependene}
\begin{figure}[h]
    \centering
        \includegraphics[width=0.9\columnwidth]{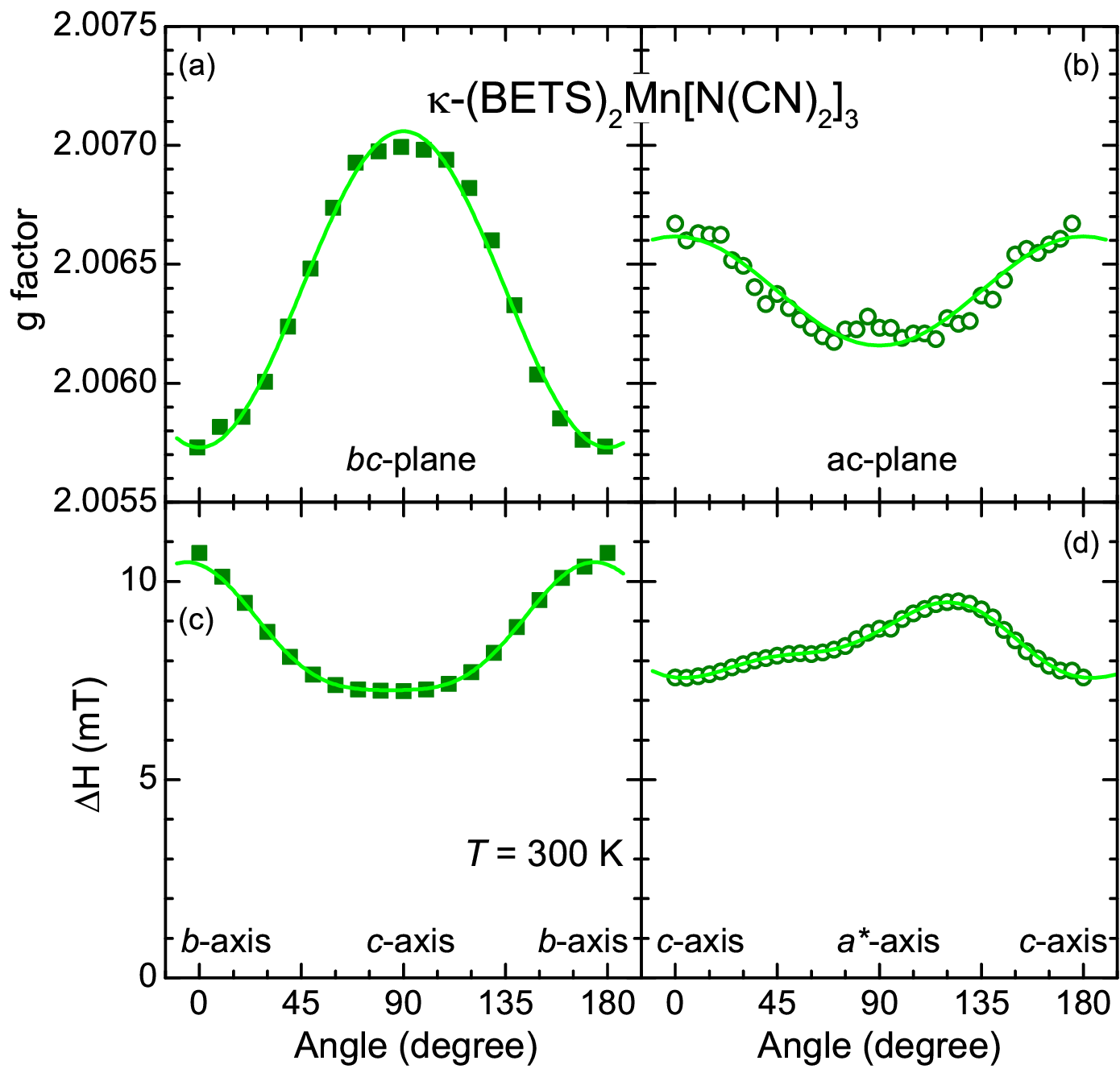}
        \caption{Angular dependence of the room-temperature ESR parameters of \kMn\ when rotated in the magnetic field.
        (a)~$g$-factor anisotropy within the $bc$-plane; and (b)~when rotating the field out of the $bc$-plane. Here $\theta=90^{\circ}$ corresponds to the $a^*$-direction, normal to the plane. The lines represent fits by Eq.~(\ref{eq:cosine2}). (c)~Angular dependence of the linewidth $\Delta H$ for in-plane rotation. (d)~Out-of-plane anisotropy of the linewidth.
        The solid lines are fits to the linewidth $\Delta H(\theta)$ taking the spin-phonon and anisotropic Zeeman interaction into account. See Fig.~\ref{fig:fit} below and the detailed discussion in Sec.~\ref{sec:discussangle}.}
    \label{fig:ESRanisotropy}
\end{figure}
The room-temperature angular dependence of the in-plane $g$-factor displayed in
Fig.~\ref{fig:ESRanisotropy}(a) follows a simple cosine behavior
\begin{equation}
\delta g(\theta) \propto \cos^2\theta \quad ,
\label{eq:cosine2}
\end{equation}
with the lowest value of $g_2=2.0057$ recorded along the $b$-axis and the largest value $g_3=2.0070$ obtained for $H\parallel c$, in accord to Fig.~\ref{fig:ESR1}(b).
Slight deviations observed close to the $c$-axis may be caused by exchange interaction; this becomes more prominent in the behavior of the linewidth $\Delta H(\theta)$. Although the quality of the data is limited due to a low signal-to-noise ratio, field perturbation and slight misalignment, the $g$-factor can be described by a $\cos^2\theta$ law when rotated out of the crystal plane, as presented in Fig.~\ref{fig:ESRanisotropy}(b).\footnote{Due to the shape of the crystal, the measurement in $bc$-plane is more reliable than in $ac$-plane. The crystal axes were determined by optical methods \cite{Schmidt24}, but there remained some error in aligning the $b$-axis parallel to the rotation axis. For that reason, the two runs arrive at different $g$-values for $H\parallel c$.}
Most importantly in the present context, this behavior does not change when cooled down to $T=2$~K.

In contrast to the $g$-factor, the angular dependence of the linewidth  is more complex, requiring a thorough  discussion, cf.\ Sec.~\ref{sec:discussion}. From a first look at the room-temperature $\Delta H(\theta)$
we already see that the angular dependence cannot be fitted by a simple dipolar interaction, suggested by spin-phonon interaction,
but requires higher-order terms \cite{Riedel75}:
\begin{equation}
\Delta H (\theta)= a_{\Delta H} + b_{\Delta H}\left(3\cos^2\theta -1\right) + c_{\Delta H} \left(3\cos^2\theta -1\right)^2
\label{eq:DeltaH}
\end{equation}
with the parameters $a_{\Delta H}= 7.7$~mT, $b_{\Delta H}=0.76$~mT and $c_{\Delta H}=0.317$~mT.
For the out-of-plane configuration, the observed behavior is even more complex.
It is also interesting to compare the findings with the structure obtained from angular magnetoresistance measurements \cite{Zverev19}.
Unfortunately, the data quality is hampered by geometry since the flake-like shape affects the field distribution.
Some misalignment of the crystals contributes further uncertainty.
Nevertheless, a thorough analysis given below allows us to decompose the contributions and extract the anisotropic Zeeman interaction in addition to spin-phonon coupling.

\section{Discussion}
\label{sec:discussion}

The radical cation salt \kMn\ is composed of two subsystems, the (BETS)$_2^+$ cations and the Mn[N(CN)$_2$]$_3^-$ anions,
both contributing to the magnetic behavior with a particular temperature and angular dependence.
Ab-initio calculations suggest that at room temperature the two subsystems couple only weakly \cite{Riedl21}.
Confining ourselves to $T > T_{\rm MIT}$, we can simply add up the various contributions when analyzing their temperature and angular dependences.

\subsection{Temperature Dependence}
\label{sec:discusstemperature}
First we consider the spin susceptibility of \kMn\ plotted in Fig.~\ref{fig:ESR1}(a),
which contains contributions from the $d$-electrons in the M[N(CN)$_2$]$_3^-$ anions and the $\pi$-electrons at the (BETS)$_2^+$ cations.
The measured temperature dependence $\chi(T)$ summarizes the response of both subsystems:
\begin{equation}
\chi(T) =  \chi_d(T) + \chi_{\pi}(T) \quad .
\end {equation}
For the paramagnetic Mn$^{2+}$ spins we can simply apply the Curie law $\chi(T) = C/T$.
The (BETS)$_2^+$ dimers are arranged on a triangular lattice [see Fig.~\ref{fig:structure}(a)] with the $\pi$-electron spins antiferromagnetically coupled; thus we assume \cite{Tamura02}
\begin{equation}
\chi(T) = \frac{C k_B}{J} \frac{4x\left(1+b_2x+ \ldots + b_6x^6\right)}{1+c_2x + \ldots + c_6 x^6}\quad
\end{equation}
with $x={J}/(4k_B T)$.

\begin{figure}
    \centering
        \includegraphics[width=0.7\columnwidth]{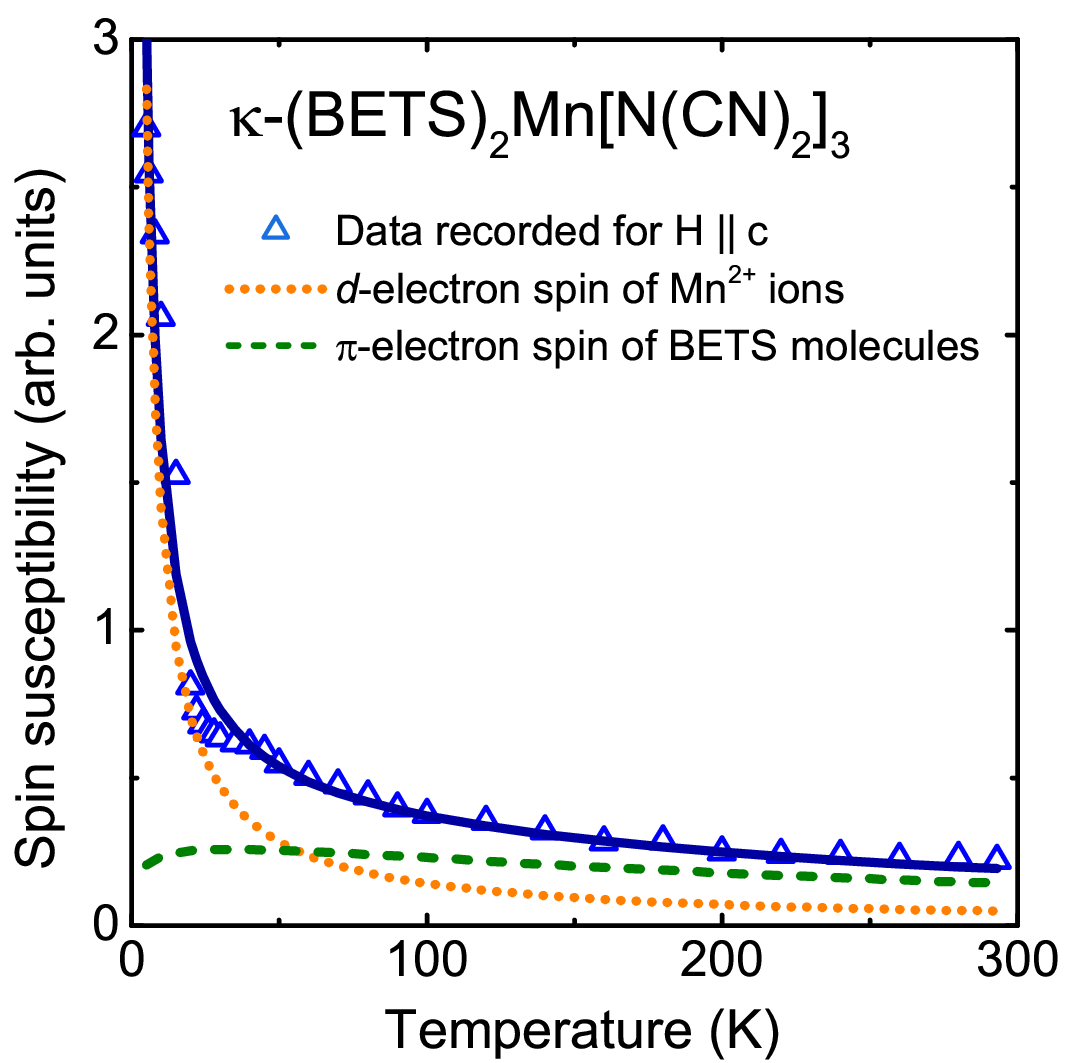}
        \caption{Temperature dependence of the spin susceptibility $\chi_c(T)$ of \kMn\ measured by X-band ESR spectroscopy with $H\parallel c$. The orange dotted line represents the contribution of the $d$-electron spin of the Mn$^{2+}$-ions located at the anion layers, while the green dashed line accounts for $\pi$-electron spins of the BETS cations. The solid blue line corresponds to the sum of both. The kink observed around $T_{\rm MI}$ indicates the rearrangement of the $\pi$-spins.}
    \label{fig:Curie}
\end{figure}
The corresponding terms are plotted in Fig.~\ref{fig:Curie} for the measurement along the $c$-axis as an example.
At room temperature, the contribution of the (BETS)$_2^+$ dimers are more pronounced;
and little changes are expected down to about 100~K. For low temperatures ($T < 50$~K), however, the Mn$~^{2+}$ spins dominate the temperature dependence completely.
The best description is obtained with
\begin{equation}
\frac{\chi_{\pi}(T)}{\chi_d(T)} \approx \frac{3}{1} \quad .
\end{equation}
Although the ratio is subject to great uncertainty, we can use it to get the combined $g$-tensor.
Obviously our simple model of adding two subsystem does not describe the kink in $\chi(T)$ observed around \TMI\ satisfactorily, evidencing the change in the magnetic properties discussed previously \cite{Vyaselev11b}.

In Fig.~\ref{fig:ESR1}(b) the temperature evolution of the $g$-factor is plotted. Within the plane ($H\parallel b$ and $H\parallel c$) a pronounced kink is observed around $T_{\rm MI} = 21$~K, which we ascribe to an increase in magnetic coupling between the $d$- and $\pi$-subsystems upon cooling. The interaction between the two magnetic subsystems is strongly visible when $H$ is aligned orthogonal to the $bc$-plane.

\subsection{Angular Dependence}
\label{sec:discussangle}

\begin{figure}[h]
    \centering
       \includegraphics[width=\columnwidth]{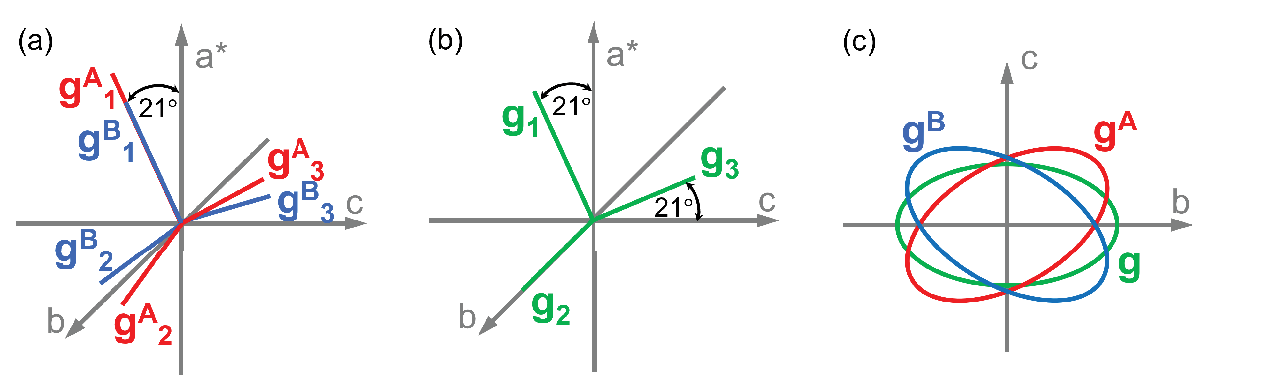}
        \caption{Relation between the principal axes of the $g$-tensor and crystal axes.  (a)~The $g$ values of the two subsystems $g^A$ and $g^B$ are described by tensors with the principal axes $g^A_1$, $g^A_2$, $g^A_3$ and $g^B_1$, $g^B_2$, $g^B_3$, respectively. The tilt of the molecules with respect to the $a^{*}$- direction causes the $g_1$ components to be rotated by $21^{\circ}$.
        The components $g^{A}_2$ and $g^{B}_2$ are slightly rotated with respect to the crystallographic $b$-axis; accordingly the $g^A_3$ and $g^B_3$-axes are slightly off.
        (b)~Due to the strong interaction of the $\pi$-electron spins
        due to exchange narrowing effects,
        the principal axes of the resulting $g^{\pi}$-tensor are averages of both components. This implies that the $g_2$-axis coincides with the crystallographic $b$-axis, while the
        principal $g_1$ and $g_3$-axes are tilted by $21^{\circ}$ with respect to the $a^{*}$ and $c$-directions. (c)~The projection of the $g$-tensors onto the $bc$-plane shows the subsystem of the A- and B-dimers by red and blue ellipses, as well as the resulting $g$-tensor of the $\pi$-electron system.}
    \label{fig:sinus1}
\end{figure}

In order to interpret the remarkable anisotropy observed in all three ESR quantities (Fig.~\ref{fig:ESR1}), let us start with the $\pi$-electron system of the organic molecules. The BETS layers are composed of the two distinct dimers, A and B,
depicted in Fig.~\ref{fig:structure}(a) by red and blue boxes. As illustrated in Fig.~\ref{fig:structure}(c), the BETS molecules do not stand normal to the $bc$-plane but are tilted by $\pm 21^{\circ}$ with respect to the $a^*$-axis. Consequently, the magnetic axes do not align with the crystallographic axes.
In Fig.~\ref{fig:sinus1}(a) the directions of the $g$-tensors are given for the A and B type (BETS)$_2^+$ dimers.
According to the molecular tilt, for both types the $g_1$-axes are consistently rotated with respect
to the $a^*$-direction by $21^{\circ}$.
The other principal axes do not coincide with each other and do not align with the crystallographic directions.
Experimentally, however, only a single $g$-value is observed [Fig.~\ref{fig:sinus1}(b)], since both contributions
add to a single $g$-tensor due to the exchange narrowing effect. The $g$-value of the entire $\pi$-electron system is thus given by the weighted average
\begin{equation}
g^{\pi}(T) = \frac{\chi^A(T)g^A(T) + \chi^B(T)g^B(T)}{\chi^A(T)+\chi^B(T)} \quad .
\label{eq:ABcoupling}
\end{equation}
Thorough studies of the vibrational features have indicated a slightly different charge residing on the (BETS)$_2^+$ dimers A and B, respectively \cite{Schmidt24}. However, the minuscule charge disproportionation of $2\delta_{\rho}= 0.02e$ inferred from infrared spectroscopy \cite{Schmidt24} is not sufficient to cause significant changes in the symmetry in the $g$-factor. At this point, we can simply assume equal magnetic strength of both dimers $\chi^A(T) = \chi^B(T)$.
The resulting angular dependence in the $bc$-plane is displayed in Fig.~\ref{fig:sinus1}(c).
While $g_2$ coincides with the crystallographic $b$-axis, the principal axes $g_1$ and $g_3$ are rotated by $21^{\circ}$, as depicted in Fig.~\ref{fig:sinus1}(b).

\begin{figure}[h]
    \centering
       \includegraphics[width=\columnwidth]{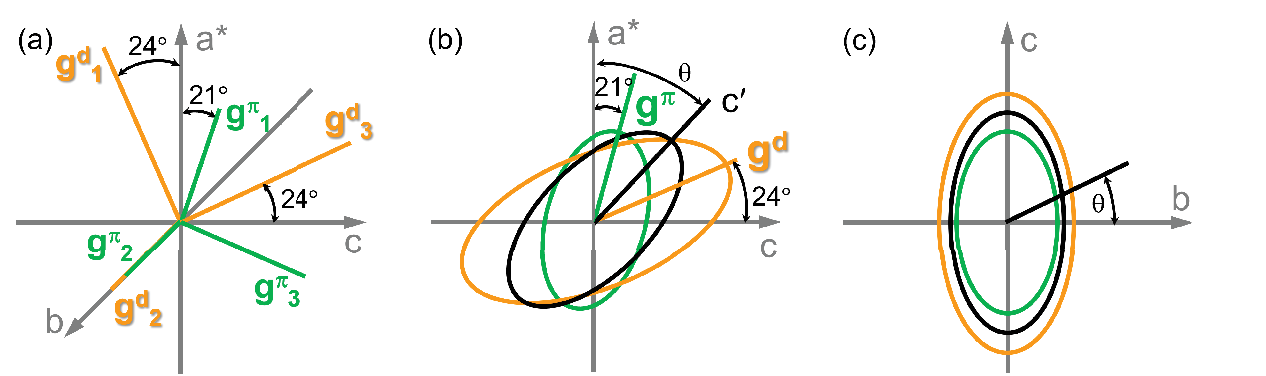}
        \caption{Relation between the principal axes of the $g$-tensors for the $\pi$-electron spins of the BETS cations and the spins of the anions, mainly determined by the Mn$^{2+}$. (a)~The principal axes of the $g^{\pi}$-tensor are shown in green; for the $d$-electrons in orange. The $g^{\pi}_1$ component is rotated by $21^{\circ}$ off the $a^*$-axis. The $g^{d}_1$-tensor is rotated by $24^{\circ}$.
        (b)~The projection of the $g$-tensors onto the $a^*c$-plane. The observed $g$-tensor is given by black. It is rotated by an angle $\theta$. (c)~The projection of the $g$-tensors onto the $bc$-plane reveals that directions of the principal axes have not changed. }
    \label{fig:sinus2}
\end{figure}

In the next step, we have to account for the Mn$^{2+}$ ions in the anion layers, resulting in an interaction of  $\pi$- and $d$-electron spins. The finally observed $g$-value is the combination of both cation ($\pi$-electrons) and anion ($d$-electrons) subsystems:
\begin{equation}
g(\theta) = \frac{\chi^{\pi}g^{\pi}(\theta) + \chi^d g^d(\theta)}{\chi^{\pi}+\chi^d}
\label{eq:pidcoupling}
\end{equation}
because the two tensors compose a new tensor describing the resulting local field.
Since the $g$-tensor of the BETS molecules does not vary among comparable materials,
we can use the previously reported data \cite{WilliamsBook,Oshima12,Lee18};
here we take: $g^{\pi}_1 = 2.008$, $g^{\pi}_2 = 2.004$, $g^{\pi}_3 = 2.006$. For the $g$-tensor of the $d$-electron spins the principal axes are obtained from the torque experiments \cite{Vyaselev17}, since at high fields, the Mn$^{2+}$ ions dominate the magnetic response completely. We use $g^{d}_1 = 2.00034$, $g^{d}_2 = 2.0109$ and $g^{d}_3 = 2.0123$. The observed $g$-factor in the $a^*c$-plane is an ellipse with the maximum value along a direction $c^{\prime}$ rotated $24^{\circ}$ off the crystallographic $c$-direction and the smallest $g$-value along the $a$-direction.
As illustrated in Fig.~\ref{fig:sinus2}, in the $bc$ plane the axes do not change. When fitting $g_b$ we obtain exactly the principal value, but in the perpendicular direction we observe $g^*_c = \left(g_a^2 \sin^2\{24^{\circ}\} + g_c^2 \cos^2\{24^{\circ}\}\right)^{1/2}$.

\begin{figure}[h]
    \centering
        \includegraphics[width=\columnwidth]{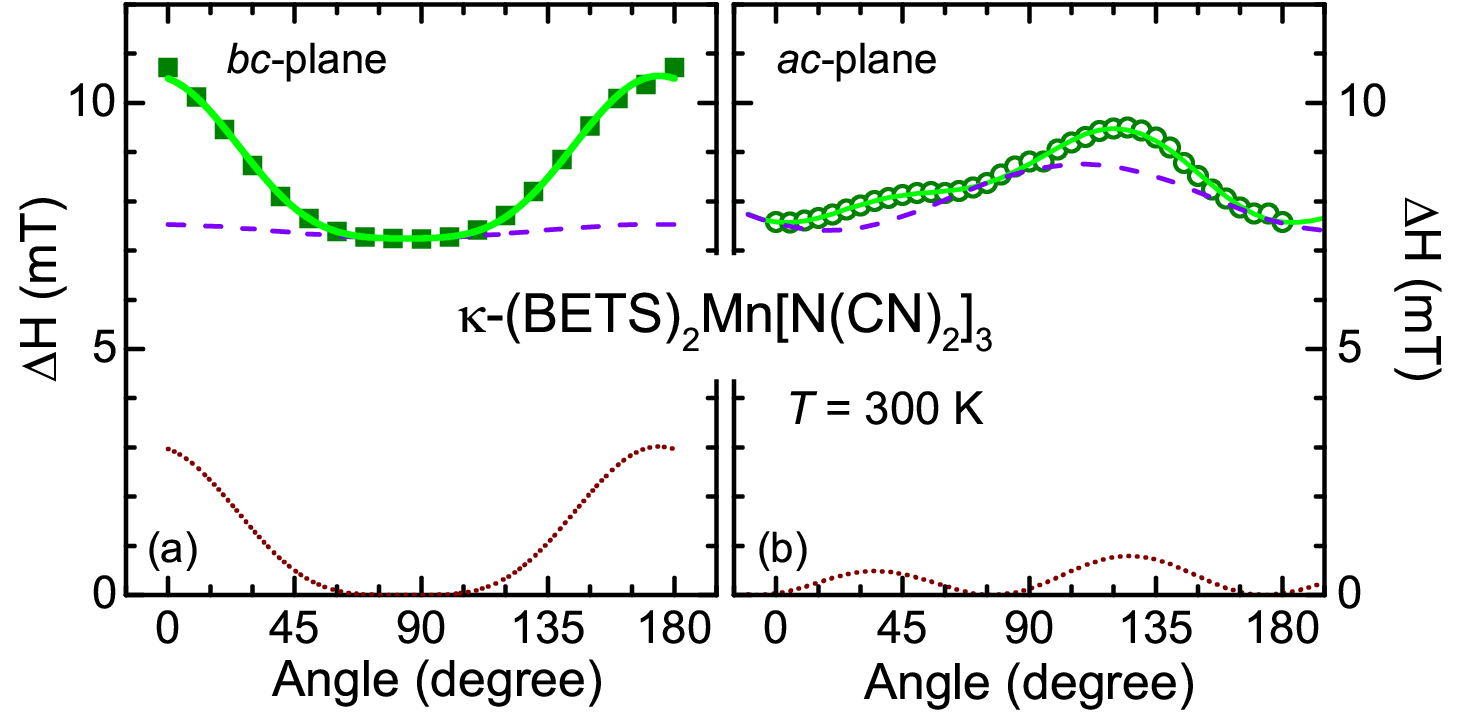}
        \caption{(a) Fit of the linewidth $\Delta H(\theta)$ of \kMn\ when rotating the magnetic field in the $bc$-plane. The dashed violet line corresponds to the spin-phonon interaction $\Delta H_{sp}(\theta)$ described by Eq.~(\ref{eq:DeltaHsp}); the dotted curve represents contributions from the anisotropic Zeeman interaction given in Eq.~(\ref{eq:DeltaHAZ}). The solid green line corresponds to the sum of both. (b) The angular dependence of $\Delta H(\theta)$ in the $ac$-plane. The fit uses Eq.~(\ref{eq:DHanisotropic}) where the first part describes the spin-phonon term (dashed violet line) and the second the anisotropic Zeeman interaction (dotted red line).}
    \label{fig:fit}
\end{figure}

While the $g$-value is a static quantity and the independent contributions simply add up according to Eq.~(\ref{eq:pidcoupling}), the linewidth reflects dynamic relaxation processes that makes it more sensitive to microscopic coupling.
For understanding the angular dependence of the ESR linewidth $\Delta H(\theta)$, two components are particularly important:
the anisotropic Zeeman interaction $\Delta H_{AZ}(\theta)$ and the spin-phonon coupling $\Delta H_{sp}(\theta)$.
The latter one has the simple form \cite{BenciniGatteschi90,Dumm00}
\begin{equation}
\Delta H_{sp}(\theta) = \left[\Delta H^2(c^{\prime})\sin^2\theta + \Delta H^2(b^{\prime})\cos^2\theta \right]^{1/2} \quad ,
\label{eq:DeltaHsp}
\end{equation}
summarizing the contributions along the principal directions $b^{\prime}$ and $c^{\prime}$.
The anisotropic Zeeman interaction is given by \cite{Pilawa97}
\begin{equation}
\Delta H_{AZ} \approx \frac{g_e (\mu_B H_0)^2}{k_B \left|J^{\prime}\right|} \sqrt{\frac{\pi}{8}} \left|\frac{\delta g}{g_e}\right|^2 \quad ;
\label{eq:DeltaHAZT}
\end{equation}
hence, the angular dependence  follows the $g$-factor anisotropy:
\begin{equation}
\Delta H_{AZ}(\theta)= c_{\Delta H_{AZ}} \delta g^2 = c_{\Delta H_{AZ}} \left[g^{\pi}(\theta) - g^{d}(\theta)\right]^2 \quad .
\label{eq:DeltaHAZ}
\end{equation}
With Equation~\ref{eq:cosine2} we arrive at the behavior of $\Delta H(\theta)$ for rotation in the $bc$-plane plotted in Fig.~\ref{fig:fit}(a).

For the out-of-plane rotation, the spin-phonon contribution $\Delta H_{sp}$ can easily be identified in Fig.~\ref{fig:ESRanisotropy}(d); unfortunately, the lineshape and quality of the data does not allow to extract the anisotropic Zeeman term with a confidence comparable to the in-plane measurements. The observed asymmetric angular dependence results from the  non-collinear $g$-tensor principal axes of these two spins, the $\pi$-electron subsystem and the $d$-electron subsystem \cite{Vyaselev17}. We fit the data by
\begin{eqnarray}
\Delta H (\theta) & = &  \left[(\Delta H_{sp1})^2 \sin^2\{\xi \} + (\Delta H_{sp3})^2 \cos^2\{\xi \}\right]^{1/2} \nonumber \\
& + &  c_{\Delta H_{AZ}}  \left[\left( (g^{\pi}_1)^2 \sin^2\{\phi \} + (g^{\pi}_3)^2 \cos^2\{\phi \}\right)^{1/2}-
\right. \nonumber \\
& & \left.\hspace*{10mm} \left((g^{d}_1)^2 \sin^2\{\psi \} + (g^d_3)^2 \cos^2\{\psi \}\right)^{1/2} \right]^{2} \quad .
\label{eq:DHanisotropic}
\end{eqnarray}
The various $g$-values have already been determined above: $g^{\pi}_1 = 2.006$, $g^{\pi}_3 = 2.008$
and $g^{d}_1 = 2.00034$, $g^{d}_3 = 2.0123$. From the fit we can define the orientation of the $g$-tensor as $\phi = \theta + 20^{\circ}$, $\psi = \theta - 51.4^{\circ}$, $\xi = \theta - 18.5^{\circ}$ and the interaction strength
$\Delta H_{sp1} = 8.758$~mT and $\Delta H_{sp3} = 7.40$~mT.
The dashed violet curve in Fig.~\ref{fig:fit}(b) represents the angular dependence of the spin-phonon contribution to the linewidth.
Since the description of the data is not satisfactory, an extra term is required, originating from antisotropic Zeeman interaction, shown by the dotted red line. It is given by the second term of Eq.~(\ref{eq:DHanisotropic}), where the
strength of the interaction $c_{\Delta H_{AZ}}=20540$~mT \footnote{We note that $c_{\Delta H_{AZ}}$ is just a fit parameter, and does not correspond to a real field. The change in linewidth due to anisotropic Zeeman interaction is of the order 10~mT because
in Eqs.~(\ref{eq:DeltaHAZ}) and (\ref{eq:DHanisotropic}) $c_{\Delta H_{AZ}}$ is multiplied by $(\delta g)^2 \approx 10^{-4}$.}, according to Eq.~(\ref{eq:DeltaHAZ}).

The dotted line in Fig.~\ref{fig:fit}(b) exhibits a $90^{\circ}$ rotational periodicity,
very much distinct from the angular dependence of all other ESR parameters, for instance the $g$-factor. It is also different from $\Delta H_{AZ}(\theta)$ when rotating in the $bc$-plane.
A similar behavior was previously identified in quasi-one-dimensional salts, such as (TMTTF)$_2$SbF$_6$,  when cooled below the charge-order transition \cite{Yasin12b,Dressel12}.
It is a clear indication of anisotropic Zeeman interaction and taken as evidence of two magnetically distinct entities in \kMn. In other words, the red and blue dimers sketched in Fig.~\ref{fig:structure}(a) do have different orientations of the magnetic moments of the $\pi$-electrons. The observation is in accord with the theoretical analysis of Riedl {\it et al.} \cite{Riedl21} and indications of a weak charge order among the dimers provided by vibrational spectroscopy \cite{Schmidt24}.
This pattern may exist down to low temperatures, but then the $d$-electron system strongly dominates (cf.~Fig.~\ref{fig:Curie}), masking the Zeeman interaction. In this case, we obtain trivial results where the angular dependence follows a simple cosine behavior.

Future ESR measurements at higher frequencies (Q-band or W-band) would be advantageous since the enhancement of the linewidth in the diagonal direction becomes more pronounced as the frequency $f$ increases: $\Delta H_{AZ}(f) = A +Bf^2$ \cite{BenciniGatteschi90,Yasin12b,Dressel12}. This way, the temperature evolution could be traced down to lower temperatures where  the effect of the $\pi$-$d$-coupling grows and finally dominates.

The low symmetry and complex magnetic structure suggests \kMn\ as a candidate of altermagnetism \cite{Smejkal22a,Smejkal22b}.
Recently, the non-linear field dependence of the Kerr angle observed in \kCl\ was interpreted as indication of altermagnetic ordering at low temperatures \cite{Naka20,Iguchi25}.
In this compound the BEDT-TTF molecules form dimers rather similar to \kMn, and the magnetic ordering at low temperatures has long been puzzling \cite{Welp92,Miyagawa95,Kubota96,Pinteric99,Antal09,Yasin11}.
Detailed magnetization measurements reveal a canted antiferromagnetic order with a pronounced field dependence. The observations are explained by a strong intralayer antiferromagnetic interaction, moderately weak Dzyaloshinskii–Moriya interaction, and a very weak interlayer ferromagnetic interaction \cite{Ishikawa18}. In a theoretical comparison of these compounds Riedl {\it et al.} suggested a two-sublattice N{\'e}el order in \kCl, while due the difference in interaction and magnetic frustration \kMn\ should exhibit a four-sublattice coplanar spin vortex crystal \cite{Riedl21}.
The presence of strong $\pi$-$d$ interaction may pave the ground for some unconventional altermagnetism in \kMn.
The angular dependence of the $g$-value reflects the static spin state only, with no possibility to identify altermagnetic ordering.
The relaxation processes resulting in the ESR linewidth $\Delta H$,
however, reflect the dynamic spin response where spin–spin interactions manifest in a non-linear way \footnote{In \kCl\ the molecules in adjacent layers are tilted in opposite directions, which makes it difficult to observe anisotropic Zeeman interactions \cite{Kubota96,Antal09,Yasin11}.
Nevertheless, detailed ESR studies of the angular dependent linewidth might reveal interesting information on the dynamic spin response. In general, angular dependent ESR experiments could be a very useful method for investigating other altermagnets.}.
The different symmetries of static altermagnetic distortion and dynamic optical response have been discussed in detail for the magneto-optical investigations of \kCl\ \cite{Iguchi25}.
While we cannot make a final conclusion here based on our ESR results, we suggest temperature and field dependent measurements of anomalous and spin Hall effects.

\section{Conclusions}

Our comprehensive X-band experiments on \kMn\ monocrystals reveal a single ESR line that exhibits  interesting temperature and angular dependences due to the subtle interplay of two magnetic subsystems: from the $\pi$-electrons of the (BETS)$_2^+$ dimers and the $d$-electrons of the Mn$^{2+}$. The temperature-dependent spin susceptibility can be well described by the antiferromagnetically arranged $\pi$-electron spins and the paramagnetic behavior of the $d$-electron spins. The latter one strongly governs the low-temperature behavior, and the signal of the (BETS)$_2^+$ dimers disappears.  This results in a cosine behavior of the $g$-factor and linewidth when the field is rotated within the plane.

At elevated temperatures, however, the $\pi$-$d$-coupling is weak. Here we can identify contributions of anisotropic Zeeman interaction with its peculiar angular dependence.
The description of the angular dependence of the linewidth $\Delta H$ within the $bc$- and $ac$-planes, where the $ac$-rotational periodicity doubles, indicates two magnetically distinct entities. Our findings provide clear experimental evidence that \kMn\ consists of two BETS chains with slightly different magnetic properties. In order to clarify the presence of some unconventional altermagnetic order further experiments at different fields          and temperatures are required.


\acknowledgments
We thank Hans-Albrecht Krug von Nidda for his illuminating comments,
Gabriele Untereiner and Sudip Pal for their technical help.
This research was funded by the Deutsche Forschungsgemeinschaft (DFG) grant numbers DR228/68-1 and SCHE1080/10-1.
We also acknowledge the support of the HLD at HZDR, member of the European
Magnetic Field Laboratory (EMFL).

%
\end{document}